\documentclass{emulateapj}
\slugcomment{Version: \today}

\shorttitle{V1647~Orionis, reinvigorated}
\shortauthors{Aspin et al.}

\begin{document}

\title{V1647~ORIONIS: REINVIGORATED ACCRETION AND THE RE-APPEARANCE OF MCNEIL'S NEBULA}

\author{
Colin~Aspin\altaffilmark{1}, 
Bo~Reipurth\altaffilmark{1}, 
Tracy~L.~Beck\altaffilmark{2},
Greg Aldering\altaffilmark{3},
Ryan~L.~Doering\altaffilmark{4}
Heidi~B.~Hammel\altaffilmark{5,9},
David~K.~Lynch\altaffilmark{6,9}, 
Margaret Meixner\altaffilmark{2}
Emmanuel Pecontal\altaffilmark{7},
Ray~W.~Russell\altaffilmark{6,9},
Michael~L.~Sitko\altaffilmark{8,9}, 
Rollin~C.~Thomas\altaffilmark{3},
Vivian U\altaffilmark{1}}

\altaffiltext{1}{Institute for Astronomy, University of Hawaii, 640 N. A'ohoku Place, Hilo, HI 96720. {\it E-mail: caa@ifa.hawaii.edu}}

\altaffiltext{2}{Space Telescope Science Institute, 3700 San Martin Drive, Baltimore, MD 2
1218.}

\altaffiltext{3}{Lawrence Berkeley Lab, Physics Div., MS-50/232, One Cyclotron Rd., Berkeley, CA 94720, USA}

\altaffiltext{4}{Department of Physics and Astronomy, Valparaiso University, Valparaiso, IN 46383.}

\altaffiltext{5}{Space Science Institute, 4750 Walnut Street, Suite 205, Boulder, CO 80301.}

\altaffiltext{6}{The Aerospace Corporation, Los Angeles, CA 90009.}

\altaffiltext{7}{Observatoire de Lyon, 9 Avenue Charles-Andre, 69561 Saint-Genis-Laval Cedex, France}

\altaffiltext{8}{Department of Physics, University of Cincinnati, Cincinnati OH 45221.}

\altaffiltext{9}{Visiting Astronomer, NASA Infrared Telescope Facility, operated by the University of Hawaii under contract with the National Aeronautics and Space Administration.}

\begin{abstract} 
  In late 2003, the young eruptive variable star V1647~Orionis
  optically brightened by over 5 magnitudes, stayed bright for around
  26 months, and then decline to its pre-outburst level.  In August
  2008 the star was reported to have unexpectedly brightened yet again
  and we herein present the first detailed observations of this new
  outburst.  Photometrically, the star is now as bright as it ever was
  following the 2003 eruption.  Spectroscopically, a pronounced
  P~Cygni profile is again seen in H$\alpha$ with an absorption trough
  extending to --700~km~s$^{-1}$.  In the near-infrared, the spectrum
  now possesses very weak CO overtone bandhead absorption in contrast
  to the strong bandhead emission seen soon after the 2003 event.
  Water vapor absorption is also much stronger than previously seen.
  We discuss the current outburst below and relate it to the earlier
  event.
\end{abstract}


\section{INTRODUCTION}
%########################################################################################
% Figure
\begin{figure*}[tb] 
\centering
\includegraphics*[angle=0,scale=0.8]{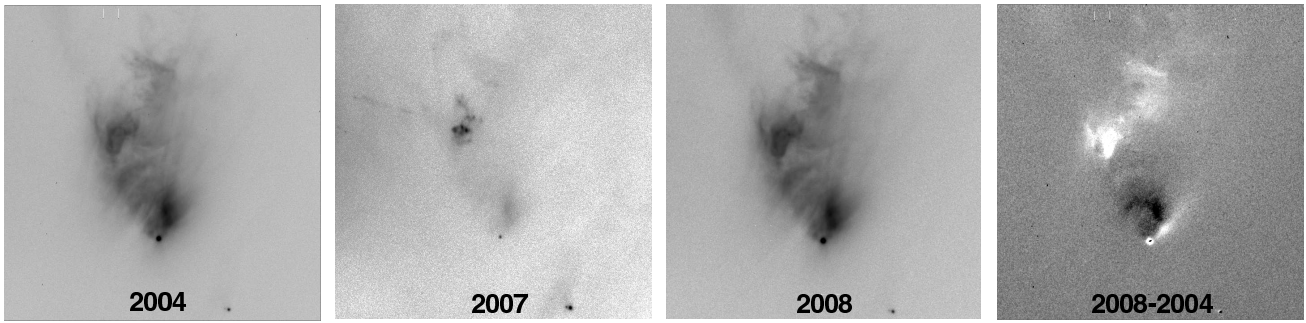} 
\caption{The region immediately surrounding V1647~Orionis and McNeil's Nebula at three different epochs.   The left image was taken in an SDSS r' filter in 2004, around one year after the 2003 outburst.  The middle-left image was taken in a standard Johnson R filter in 2007, around 10 months after V1647~Ori had returned to its pre-outburst optical brightness.  The middle-right image was taken in the same R filter in 2008, soon after the recent brightening of the star and the re-appearance of McNeil' Nebula.  The right image is a difference image, 2008--2004.  In this image, white is more flux in 2008, black is more flux in 2004.  We note that V1647~Orionis is the point-like object at the apex of the monopolar reflection nebula extending northward.  The images were scaled to result in the cancellation of the faint binary star at the bottom of the image. 
\label{3images}}
\end{figure*}
%\clearpage
%########################################################################################
When V1647~Orionis went into outburst in the fall of 2003, it brightly
illuminated a monopolar reflection nebula subsequently designated
McNeil's Nebula (McNeil 2004).  The young star itself increased in
optical and near-IR (NIR, K-band) brightness by around 5 and 3
magnitudes, respectively, and sparked a world-wide effort to study the
details of this rare type of eruption.  To date, this has resulted in
over 50 research papers which relate the characteristics of the star
and nebula from the time of its detection to the time when it had
faded to its pre-outburst optical brightness (in February 2006).  The
reader is referred to the most recent publication on V1647~Ori,
(Aspin, Beck, \& Reipurth 2008, ABR08), for a comprehensive list of
previous work.

It has been speculated that V1647~Ori may be a new example of the
class of young variable stars designated ``FUor'' after the class
progenitor, FU~Orionis (Ambartsumian 1971).  Only about ten so-called
``classical'' FUors, (those that have been directly observed to
optically brighten over a period of a few weeks to a few months) have
so far been found.  A comprehensive review of the properties of FUors
was given by Hartmann \& Kenyon (1996), however, it is important to
relate here the main characteristics defining this group.  {\it i)} an
observed rise in optical brightness of around 5--6 magnitudes, {\it
  ii)} an optical spectrum indicative of a G- to F-type supergiant
star, {\it iii)} a NIR K-band spectrum typical of a M-type giant, {\it
  iv)} the star remains in an elevated state for many years, even
decades -- FU~Ori has declined little in over the 70~years since its
discovery (Herbig (1966, 1977), {\it v)} an H$\alpha$ line in emission
with an associated blue-shifted absorption component (i.e. a P~Cygni
profile), and {\it vi)} an associated, generally curving, reflection
nebula.  What has been gleamed with regard to V1647~Ori, however, is
somewhat contradictory to these definitive characteristics in that it
had a relatively short outburst period (approximately 26 months), its
optical and NIR outburst spectra possessed strong atomic and molecular
emission features, and the eruption was shown to be repetitive --
Aspin et al. (2006) described an earlier event commencing in 1966.
Such properties have more in common with those observed in the
``EXor'' class of eruptive variables, named after their class
progenitor, EX~Lupi (Herbig 1989).

% However, recent high spectral resolution NIR observations have
% demonstrated a remarkable spectroscopic correspondence between
% V1647~Ori and FUors (Aspin, Greene, \& Reipurth 2009, henceforth
% AGR09).  It is therefore unclear whether V1647~Ori is a FUor or an
% EXor, or, as suggested in Brittain et al. (2007) and AGR09, part of
% a continuum of outbursts characteristics with FUors and EXors
% perhaps being the extremes.

It was expected that the next eruption of V1647~Ori would be $\sim$37
years after the 2003 event, assuming the timescale between the 1966
and 2003 events was typical.  However, in late-August 2008, some 18
months after the fading of the 2003 outburst, V1647~Ori was observed
to brighten by around 5 magnitudes yet again and McNeil's Nebula
became brightly visible (Itagaki et al. 2008).  This was confirmed by
Aspin (2008) who showed that V1647~Ori had brightened in the red
optical ($\sim$6500~\AA) from r'=23.3 in March 2006 to R=17.3 in
late-August 2008.  Here we present new optical and NIR observations of
V1647~Ori from late-August to early-September 2008.  In $\S~$2 we
briefly describe the observations and data reduction.  In $\S~$3 we
present the results, while in $\S~4$ we discuss the current event in
relation to the 2003 outburst.

%########################################################################################
% Table Obslog
%\begin{deluxetable}{ccccc}
%\tabletypesize{\scriptsize}
%\tablecaption{Observation Log\label{obslog}}
%\tablehead{
%\colhead{UT} & \colhead{Telescope} & \colhead{Inst.} & \colhead{Exp.Time} & %\colhead{S:N\tablenotemark{a}} \\
%\colhead{Date} & \colhead{Used} & \colhead{Used} & \colhead{(secs)} & %\colhead{}}
%\startdata
%2004/09/03 & Gemini-N & GMOS   & 30 & -- \\
%2007/12/22 & UH~2.2m  & TEK2K  & 120 & -- \\
%2008/08/31 & UH~2.2m  & SNIFS  & 900 & 20--120 \\
%2008/09/01 & UH~2.2m  & TEK2K  & 270 & -- \\
%2008/09/04 & IRTF     & SpeX   & 960 & 100--200 \\
%2008/09/13 & Gemini-S & T-ReCS & 120/120\tablenotemark{b} & -- \\ 
%2008/09/16 & WIYN     & WHIRC  & 20/20/12\tablenotemark{d} & -- \\
%2008/09/22 & Gemini-S & GMOS   & 240/240/120/80\tablenotemark{c} & -- \\ 
%2008/09/22 & Gemini-S & GMOS   & 900 & 50--200 \\ 
%\enddata
%\tablenotetext{a}{Signal to noise in spectroscopic observations.}
%\tablenotetext{b}{Total on-source exposure time using the N and Qa filters.}
%\tablenotetext{c}{Total on-source exposure time in g', r', i', and z' %filters.}
%\tablenotetext{d}{Total on-source exposure time in J, H, and Ks filters.}
%\end{deluxetable}
%\clearpage
%########################################################################################
\section{OBSERVATIONS AND DATA REDUCTION}
Below we present new data taken on four different telescopes and with
five different instruments.  Optical R-band images (see
Fig.~\ref{3images}) were obtained on the University of Hawaii (UH)
2.2~meter telescope on UT 2008 September 1 using the Tektronix
2K$\times$2K CCD camera (Tek2K).  The on-source exposure time was
120~seconds and seeing was $\sim$0$\farcs$7.
% The data was reduced using the Starlink CCDPACK software (Draper et
% al. 2006).
The images were calibrated by using a sequence of six field stars
described in Aspin, Herbig, \& Reipurth (2009, henceforth AHR09).

Optical images were also obtained on Gemini-South using GMOS (Davies
et al. 1997; Hook et al. 2004) on UT 2008 September 22.  The SDSS g',
r', i', and z' filters were used and the on-source exposure times were
240, 120, 120, and 80~seconds, respectively.  
% The data were reduced using the Gemini iraf package v1.9 GMOS
% reduction scripts and
Details of the photometric calibration, can be found AHR09. Seeing
during the observations was $\sim$0$\farcs$6.
 
Optical spectroscopy was obtained on the UH~2.2~meter telescope on UT
2008 August 31 using the dual-channel integral-field unit (IFU)
spectrograph, SNIFS (Aldering et al. 2002; Lantz et al. 2004).  The
6$''$$\times$6$''$ field-of-view of the SNIFS spectrograph is filled
with a 15$\times$15 grid of lenslet-defined spatial elements, each
sampling $\sim$0$\farcs$4.  The data was reduced by the SNIFS
reduction pipeline (Aldering et al. 2006).  The spectral resolution of
the final spectrum was R$\sim$1850 which was extracted from the IFU
cube using a wavelength-dependent PSF fitting procedure (Aldering et
al. 2006).  Wavelength calibration was performed using the SNIFS
internal calibration unit and flux calibration was performed using
observations of spectrophotometric standards.

An optical spectrum was also obtained on Gemini-South using GMOS on UT
September 22.  We used the R400 grating and a 0$\farcs$75 slit giving
an effective spectral resolution of R$\sim$1900.  Data was taken in
nod-and-shuffle mode to aid in the removal of the bright red optical
sky emission lines.  
% The data was again reduced using the Gemini iraf package v1.9 (see
% AHR09 for procedural details).

NIR spectroscopy was obtained on the NASA IRTF telescope on UT 2008
September 4 using the facility NIR spectrograph SpeX (Rayner et
al. 2003).  The cross-dispersed (XD) mode was used and observations
were acquired in both the short- and long-XD settings.  The data were
reduced using the SpeX IDL package (Cushing, Vacca, \& Rayner 2004).
Telluric correction and flux calibration were performed using the A0~V
telluric standard star HD~39985.  Conditions were clear and the seeing
stable during the target and standard observations and hence we
consider the flux calibration good to the 10\% level.

Mid-IR (MIR) imaging was obtained on Gemini-South on UT September 13,
2008 with T-ReCS (Telesco et al. 1998) using broad N'
($\lambda_c$=11.2~$\mu$m) and narrow Qa ($\lambda_c$=18.30~$\mu$m)
filters.  Conditions were clear, dry and stable throughout the
observations and photometric calibration was performed using similar
observations of the bright ``Cohen'' standard star (Cohen et al. 1999)
HD39400 (N'=1.534, Qa=1.434).  
% The images were reduced using the Gemini iraf package v1.9 and the
% ``midir'' scripts.  Specifically, the task ``mireduce'' was used in
% all cases.

Near-IR images were obtained on the WIYN 3.5-meter telescope on UT
September 16 2008.  The WHIRC NIR camera (Meixner et al. 2008) was
used with standard J, H, and Ks filters.  
%Data was reduced using standard NIR imaging techniques and 
Flux calibration was achieved using observations of the UKIRT faint
standard star FS~07 (Hunt et al. 1998).

%########################################################################################
% Figure
\begin{figure*}[t] 
\centering
\includegraphics*[angle=0,scale=1.3]{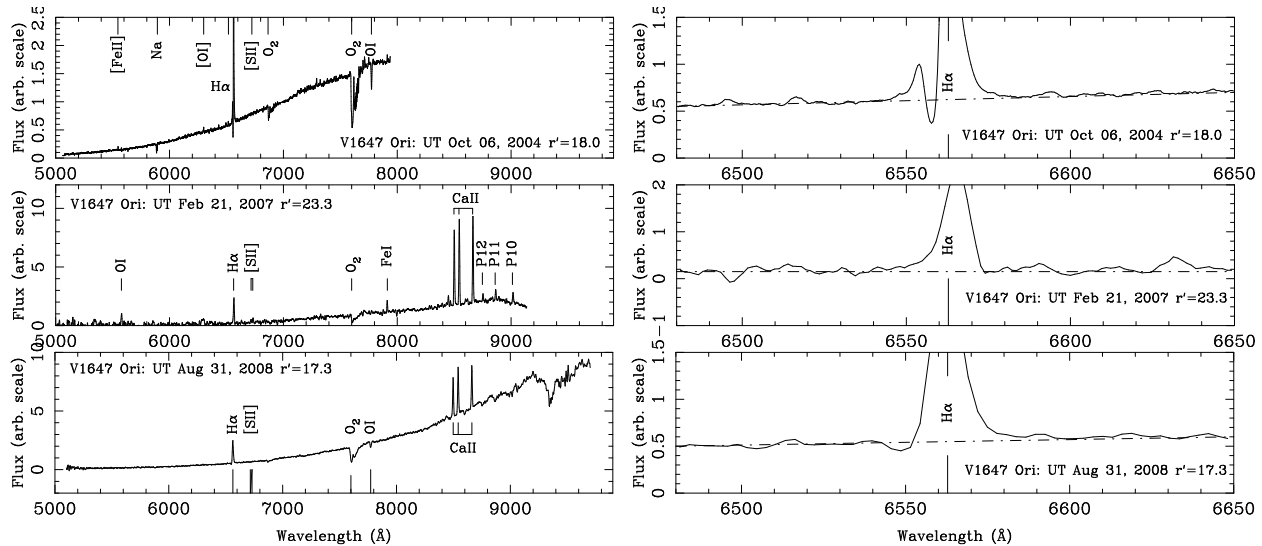} 
%\epsscale{0.45}
%\plotone{3opt.ps} 
\caption{Optical spectra of V1647~Ori from three epochs, UT August 31, 2008, February 21, 2007, and October 6, 2004.  The 2008 spectrum is from the latest outburst of the star while the 2004 spectrum was taken during the 2003--2006 event.  The 2007 spectrum was taken around one year after V1647~Ori had returned to its pre-outburst optical brightness after the 2003 eruption had subsided. Note the 2004 spectrum did not include the Ca~II lines.  The left three panels show the full spectra range observed while the right three panels focus on the region immediately around the H$\alpha$ emission feature.
\label{3opt}}
\end{figure*}
%\clearpage
%########################################################################################
\section{RESULTS}
In Fig.~\ref{3images}, we show a comparison of red optical images of
V1647~Ori from three different epochs.  The left image is from UT 2004
September 3 and was taken with an SDSS r' filter using GMOS.  This was
around 10--11 months after the 2003 outburst of V1647~Ori was first
detected (McNeil 2004) and V1647~Ori had a magnitude of r'=17.8.  The
middle-left image shows the same region of sky but was taken on UT
2006 December 22 using a Johnson R filter with the UH Tek2K camera.
At the time of these observations, V1647~Ori had been at its
pre-outburst (red optical) brightness of r'=23.3 for around 10 months.
The middle-right is our recent UH Tek2K image detailed in $\S~$2
above.  In this image, V1647~Ori has an optical brightness of R=17.3.
Comparing the left and middle-right images, we note that in both
V1647~Ori and its associated nebula are strongly visible.  However, in
the middle-left image both the star and the nebula are only faintly
visible.  The bright knot visible in the middle-left image (top and
left of center) is the Herbig-Haro object, HH~22.  We note that the
optical brightness of V1647~Ori in the 2004 and 2008 images is similar
as is the brightness and morphology of McNeil's Nebula.  The far-right
image shows the difference between the 2004 and 2008 images with white
implying that the object was brighter in 2008, and black meaning
brighter in 2004.  The images were scaled prior to subtraction to give
the same signal in the faint binary system near the bottom (and left
of center) of the images.  Clearly, there are significant differences
in brightness between the two epochs suggesting that in 2008, the
reflection nebula is illuminated in a somewhat different manner than
in 2004.  This suggests that dust obscuration close to V1647~Ori may
play a important role in defining the observed morphology of the
nebula.

Photometric observations of V1647~Ori from the optical through mid-IR
are shown in Table~\ref{phot}.  Optical and NIR colors are also given.
Within the associated uncertainities, the optical through NIR
brightness values for V1647~Ori are comparable to those measured in
February 2004 when the source was last observed to have brightened
significantly (Reipurth \& Aspin 2004).

We present our new optical spectrum of V1647~Ori in Fig.~\ref{3opt}
(bottom).  We additionally show similar spectra from October 2004
(top) and February 2007 (middle) for direct comparison.  In October
2004, V1647~Ori was some 11--12 months into its eruption and had not
yet started to decline in optical brightness.  In February 2007, the
star had been at its pre-outburst optical brightness for around 1
year.  In all three spectra, H$\alpha$ is in emission although only in
October 2004 and August 2008 did it show a blue-shifted absorption
component (a P~Cygni profile).  On both these dates, the blueshifted
H$\alpha$ absorption extended to approximately --700~km~s$^{-1}$
although in August 2008 the depth of the absorption is less than in
October 2004.  The full-width half maximum of the H$\alpha$ emission
is large with the line wings extending to around
$\pm$1000~km~s$^{-1}$.  This is somewhat larger than observed at the
start of the 2004 event (i.e. $\sim$750~km~s$^{-1}$, see AHR09 for
more details) suggesting the physical conditions in the H emission
region are somewhat more extreme.  Also present in the August 2008
spectrum are the Ca~II triplet lines in emission.  The ratios of these
lines (1.6:2:1.7) are similar to those found in February 2004
(1.5:2:1.7) implying that unlike the emission seen in February 2007
(with ratios 1.7:2:1.9), conditions in the Ca~II emission region have
become optically thick producing saturated line emission.  The Ca~II
emission lines are discussed in more detail in ABR08.  Also present in
our recent spectrum are O~I (6300~\AA) in emission, and O~I (7773~\AA)
in absorption.  Other weak features may be present but, with the
spectral resolution obtained with SNIFS, it is difficult to determine
if they are real. Included in Table~\ref{lines} are details of the
optical spectral features present.  Two lines that are conspicuously
absent are the [S~II] lines at 6717 and 6731\AA\ which were present in
February 2007.  This is likely due to the already weak lines (see
ABR08) being masked by the addition of strong continuum (accretion)
emission which was not present in the February 2007 spectrum.
Finally, we note that the SNIFS acquisition image of V1647~Ori gave a
magnitude of V=18.1.

%########################################################################################
% Table Photometry
\begin{deluxetable}{lccc}
%\tabletypesize{\scriptsize}
\tablecaption{Photometry of V1647~Ori\label{phot}}
\tablehead{
\colhead{Filter($\lambda$)} & \colhead{UT} & \colhead{2008 Flux\tablenotemark{a}} & \colhead{2004 Flux\tablenotemark{b}} \\
\colhead{($\mu$m)} & \colhead{} & \colhead{(mags)} & \colhead{(mags)}}
\startdata
g' (0.475) & 2008/09/22 & 19.86$\pm$0.12  & 20.05$\pm$0.15 \\
V (0.55)   & 2008/08/31 & 18.10$\pm$0.15  & 18.50$\pm$0.15\tablenotemark{c} \\
r' (0.63)  & 2008/09/22 & 17.53$\pm$0.07  & 17.70$\pm$0.09 \\
i' (0.78)  & 2008/09/22 & 16.03$\pm$0.06  & 15.90$\pm$0.08 \\
z' (0.925) & 2008/09/22 & 14.43$\pm$0.06  & 14.39$\pm$0.06 \\
J (1.25)   & 2008/09/16 & 10.62$\pm$0.07  & 10.85$\pm$0.10\tablenotemark{c} \\
H (1.65)   & 2008/09/16 & 8.90$\pm$0.07   &  8.90$\pm$0.10\tablenotemark{c} \\ 
Ks (2.2)   & 2008/09/16 & 7.53$\pm$0.07   &  7.60$\pm$0.10\tablenotemark{c} \\ 
N' (11.2)  & 2008/09/13 & 2.51$\pm$0.20   &  1.50$\pm$0.20 \\ 
Qa (18.3)  & 2008/09/13 & 0.62$\pm$0.20   &  -- \\ 
g'-r'      & 2008/09/22 & 2.33$\pm$0.17   & 2.35$\pm$0.12 \\ 
r'-i'      & 2008/09/22 & 1.50$\pm$0.10   & 1.80$\pm$0.09 \\ 
r'-z'      & 2008/09/22 & 1.60$\pm$0.10   & 1.51$\pm$0.08 \\ 
J-H        & 2008/09/22 & 1.72$\pm$0.10   & 1.95$\pm$0.14 \\ 
H-K        & 2008/09/22 & 1.37$\pm$0.10   & 1.30$\pm$0.14 \\ 
\enddata
\tablenotetext{a}{Data acquired during the current outburst.}
\tablenotetext{b}{Data acquired in Feb--Mar 2004.}
\tablenotetext{c}{Data from McGehee et al. (2004).}
%\tablenotetext{c}{Data from Vacca et al. (2004).}
\end{deluxetable}
%\clearpage
%########################################################################################
%########################################################################################
% Table Obslog
\begin{deluxetable}{ccccc}
\tabletypesize{\scriptsize}
\tablecaption{Spectral Features\label{lines}}
\tablehead{
\colhead{$\lambda$} & \colhead{Line} & \colhead{W$_{\lambda}$} & \colhead{Flux} & \colhead{Comment} \\
\colhead{($\mu$m)} & \colhead{ID} & \colhead{(\AA)} &\colhead{(Wm$^{-2}\mu$m$^{-1}$)} & \colhead{} \\
\colhead{} & \colhead{} & \colhead{} &\colhead{(10$^{-17}$)} & \colhead{}}
\startdata
0.6300 & O~I       & --1.1   & 0.02  & -- \\
0.6563 & H$\alpha$ & --39/+1 & 0.89  & P~Cygni\tablenotemark{a} \\
0.7773 & O~I       & +5      & --    & -- \\
0.8498 & Ca~II     & --7     & 1.26  & -- \\
0.8542 & Ca~II     & --8     & 1.60  & -- \\
0.8662 & Ca~II     & --6     & 1.35  & -- \\
1.0830 & He~I      & --2/+7  & 1.55  & P~Cygni\tablenotemark{b} \\
1.2818 & Pa$\beta$ & --6     & 8.91  & 470~km~s$^{-1}$\tablenotemark{c} \\
1.6444 & [Fe~II]   & --0.3   & 0.91  & very weak \\
2.0581 & He~I      & --1/+2  & 3.98  & P~Cygni\tablenotemark{d} \\
2.1659 & Br$\gamma$& --5     & 18.6  & 665~km~s$^{-1}$\tablenotemark{c} \\
2.2940 & CO        & +1      & --    & weak bandheads \\
4.0509 & Br$\alpha$& --10    & 35.5  & 415~km~s$^{-1}$\tablenotemark{c} \\
4.6741 & CO$_{ice}$ & +57     & --    & 830~km~s$^{-1}$\tablenotemark{c} \\
\enddata
\tablenotetext{a}{FWHM=450~km~s$^{-1}$, absorption to --950~km~s$^{-1}$, minimum at --690~km~s$^{-1}$, red tail to +830~km~s$^{-1}$.}
\tablenotetext{b}{FWHM=320~km~s$^{-1}$, absorption to --760~km~s$^{-1}$, minimum at --300~km~s$^{-1}$.}
\tablenotetext{c}{FWHM of emission component.}
\tablenotetext{d}{FWHM=400~km~s$^{-1}$, absorption to --850~km~s$^{-1}$, minimum at --290~km~s$^{-1}$.}
\end{deluxetable}
%\clearpage
%########################################################################################
% Figure
\begin{figure*}[tb] 
\centering
\includegraphics*[angle=0,scale=0.95]{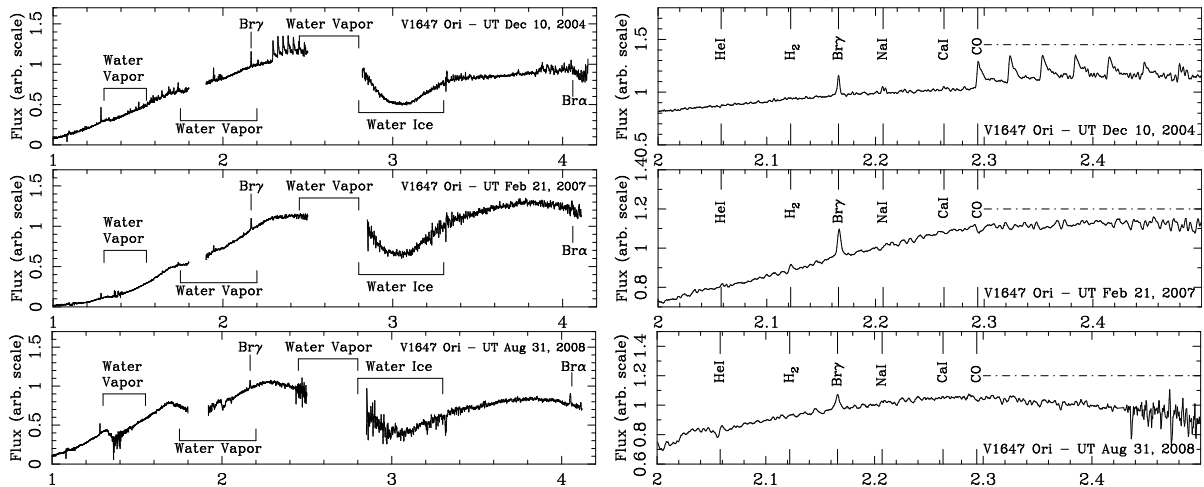} 
%\epsscale{0.45}
%\plotone{3dates.ps} 
\caption{The NIR 1 to 4.2~$\mu$m SpeX spectrum of V1647~Ori taken on UT August 31, 2008 (bottom panels).  For comparison, we also show the same wavelength range from SpeX spectra from UT February 21, 2007 (middle panels) and UT December 10, 2004 (top panels).   The main spectral features are identified.  Features present are water vapor absorption bands (in J, H, and K), water ice absorption (at 3~$\mu$m), and emission lines of Pa$\beta$ (at 1.282~$\mu$m), weak [Fe~II] (at 1.644~$\mu$m), He~I (at 2.058~$\mu$m), Br$\gamma$ (at 2.166~$\mu$m), and Br$\alpha$ (at 4.051~$\mu$m).  The right three panels zoom into the  K-band region of the three spectra.  Note the CO bandheads (at 2.294~$\mu$m) are in emission in 2004, are weakly in absorption in 2007, and perhaps very weakly in absorption in 2008.   
\label{3nir}}
\end{figure*}
%\clearpage
%########################################################################################
Fig.~\ref{3nir} shows the SpeX NIR spectrum of V1647~Ori (bottom).  As
in the optical, we also show spectra from two other epochs, in this
case, February 2007 (middle) and December 2004 (top).  Present in the
August 2008 NIR spectrum are the atomic emission lines of He~I
(1.083~$\mu$m, P~Cygni profile), Pa$\beta$ (1.282~$\mu$m), weak
[Fe~II] (1.644~$\mu$m), He~I (2.058~$\mu$m, P~Cygni profile),
Br$\gamma$ (2.166~$\mu$m), and Br~$\alpha$ (4.051~$\mu$m).  Details of
these features are given in Table~\ref{lines}.  The molecular CO
overtone bandheads longwards of 2.294~$\mu$m are weakly in absorption
in both February 2007, very weakly in absorption in August 2008, while
in October 2004 they were strongly in emission.  The water vapor
absorption bands, at 1.4, 1.9, and 2.5~$\mu$m, are strongly in
absorption, and appear much deeper in August 2008 than in either
February 2007 or December 2004.  The other broad feature seen in
August 2008 is the water ice absorption band, centered at around
3~$\mu$m.  It appears somewhat different in shape in August 2008 with
respect to the two earlier spectra with a less rounded minimum and a
more gradual increase back to the continuum on the long-wavelength
side.  Finally, the August 2008 spectrum shows the presence of the
4.674~$\mu$m CO ice absorption band (not included in the plot shown in
Fig.~\ref{3nir}).  This band was previously observed in V1647~Ori by
Vacca et al. (2004), Rettig et al. (2005), and Gibb et al. (2006).
Rettig et al. modeled the band shape using polar, and apolar CO ice
and found that a predominantly apolar CO matrix ice best fit the
February 2004 data with a temperature of $<$20~K.  Their conclusion
was that the line-of-sight CO ice had not been thermally processed and
was consistent with that found towards quiescent dark clouds and
region of low-mass star formation.  A qualitative comparison of the
band shape in our August 2008 spectrum with that in the February 2004
spectrum of Vacca et al. (2004) suggests that little has changed in
the characteristics of the CO ice absorption in the intervening four
years.

Using the Pa$\beta$ and Br$\gamma$ line fluxes, we can estimate mass
accretion luminosities (L$_{acc}$) and rates
(log(M$_\odot$~yr$^{-1}$)) in a manner similar to that presented in
ABR08.  Using the relationships defined by Muzerolle et al. (1998) and
Gullbring et al. (1998), we find that in August 2008 the line fluxes
shown in Table~\ref{lines}, resulted in values of
L$_{acc}$=1.42~L$_{\odot}$ and log(M$_\odot$~yr$^{-1}$)=--6.28 for
Pa$\beta$, and L$_{acc}$=12~L$_{\odot}$ and
log(M$_\odot$~yr$^{-1}$)=--5.34 for Br$\gamma$.  This assumes a
stellar temperature of T$_{eff}$=3800~K, a stellar luminosity of
L$_{*}$=4.9~L$_{\odot}$, and an inner radius for the accretion disk of
R$_{in}$=5~R$_{*}$ as in ABR08.  For the August 2008 fluxes, we also
assumed a visual extinction of A$_V$=6 magnitudes.  This was
determined by dereddening the observed NIR (J-H and H-K) colors of
V1647~Ori to typical T~Tauri star values as in ABR08 (see their
Figs.~13 and 14).  We note that the values of L$_{acc}$ and
log(M$_\odot$~yr$^{-1}$) derived from Pa$\beta$ and Br$\gamma$ are
inconsistent being larger from Br$\gamma$ by factors of $\sim$8 and
$\sim$9, respectively.  No value of A$_V$ resulted in similar values
from the two lines.  We can perhaps attribute this difference to the
emission in Pa$\beta$ being optically thick while in Br$\gamma$ the
emission is less so.  The ratio of the dereddened (A$_V$=6) fluxes,
Pa$\beta$/Br$\gamma$, is $\sim$1 which, from the observations and
analysis shown in Fig.~15 from Muzerolle et al. (2001), is a factor
$\sim$5$\times$ smaller than that observed in typical 'quiet'
classical T~Tauri stars.  We note that in February 2007, when
V1647~Ori was supposedly in quiescence, ABR08 derived values of
$\sim$4~L$_{\odot}$ and $\sim$--6 (using the same physical parameters
as above with A$_V$=19) from both lines, and a line ratio of $\sim$5.
Clearly, the changes observed in L$_{acc}$ and log(M$_\odot$~yr$^{-1}$)
warrant a more detailed investigation.

Our MIR imaging of V1647~Ori showed only a point source and hence, is not explicitly presented here.  However, we have extracted photometric values from these data and can compare them to earlier measurements.   We find that the integrated flux from V1647~Ori at N' and Qa is 3.8 and 6.6~Jy, respectively.   Similar observations of V1647~Ori in February 2007, when the star had faded to its pre-outburst optical brightness, gave N' and Qa fluxes of 0.23 and 0.44~Jy, respectively (ABR08).   Earlier measurements quoted by Andrews, Rothberg, \& Simon (2004) showed V1647~Ori to have a 11.3~$\mu$m flux of $\sim$10~Jy.  The latter observations were taken soon after the source was discovered by J.~McNeil, specifically, in March 2004.  Our N' flux of 3.8~Jy is a factor of 2.6$\times$ smaller than the 2004 outburst value and 16.5$\times$ larger than the quiescent phase flux.  At 18.6~$\mu$m (Qa) our flux measurement from September 2008 is 15$\times$ the flux observed in 2007.   Thus, the ratio of N/Qa flux in September 2008 is the same (within the associated uncertainities) as in February 2007 at around 0.55$\times$. \\

\section{SUMMARY}
What we can conclude about the late-August 2008 reappearence of
V1647~Ori and McNeil's nebula is that:

In the optical, V1647~Ori and McNeil's Nebula appear
photometrically and morphologically similar to when the source was
first observed in 2004 i.e. shortly after the eruption occurred.

% 2. Its optical spectrum exhibits a strong, red continuum, as it did
% in February 2004 (Reipurth \& Aspin 2004).  H$\alpha$ is strongly in
% emission with a blue-shifted absorption component -- a P~Cygni
% profile (Reipurth \& Aspin 2004; Walter et al. 2004).  The
% full-width zero intensity of the H$\alpha$ line is around
% 2000~km~s$^{-1}$ (it was $\sim$1500~km~s$^{-1}$ in February 2004)
% and the absorption component extends to 700~km~s$^{-1}$ from the
% rest wavelength (similar to that observed in February 2004).  Also
% in emission are the red Ca~II triplet lines with ratios (1.6:2:1.7)
% similar to those observed in February 2004 (1.5:2:1.7), one year
% after it had returned to its pre-outburst brightness.

In the NIR, V1647~Ori appears photometrically similar to February
2004.  Spectroscopically, unlike at the start of the 2003 event, CO
overtone emission is not observed.  Br$\gamma$ and Pa$\beta$ emission
are however present.  Strong water vapor bands are also seen.

The values of L$_{acc}$ and log(M$_\odot$~yr$^{-1}$) derived from
the dereddened Br$\gamma$ flux (12~L$_{\odot}$ and -5.3, respectively)
are of a similar order to those found soon after the 2004 outburst
(Muzerolle et al. 2005).

% 5. The MIR 10~$\mu$m flux observed is a factor of 2.6$\times$
% smaller than in February 2004 but is a factor $\sim$16$\times$
% larger than in February 2007.

Our qualitative interpretation of the results presented above (for the
current eruption, the appearance of V1647~Ori in 2007 as described by
ABR08, and for the October 2003 eruption as described by AHR09) can be
summarized as follows.  We believe that the massive accretion event
that occurred in 2003 did not actually terminate in 2006 as previously
discussed, but only declined in intensity.  This is supported by the
high-level of accretion found in February 2007 by ABR08, specifically,
log(M$_\odot$~yr$^{-1}$)$\sim$--6.  We suggest that the $\sim$6
magnitude optical fading of V1647~Ori between late 2005 and early 2006
was the result of not only a factor 10$\times$ reduction in accretion
luminosity (Muzerolle et al. 2005 derived a value of --5), but also
the reformation of dust in the immediate circumstellar environment of
the star which had been sublimated by the intense radition field from
the 2003 accretion burst.  We add further support for this speculative
interpretation by noting that the change in A$_V$ observed between
December 2004 and December 2005 (see Fig.~14 of ABR08) was of a
similar order.  In early 2008, we consider that a second burst of
accretion again sublimated circumstellar dust and resulted in the
current $\sim$6 magnitude brightening of V1647~Ori.
% Whether this reappearance is an isolated occurrence and V1647~Ori
% will now fade to quiescence, or the current period of relative
% instability will continue for some time to come, is unclear.
% However, this young eruptive variable clearly warrants further
% monitoring.

\vspace{0.3cm}

{\bf Acknowledgments} 
Based on observations obtained at the Gemini Observatory, which is operated by the Association of Universities for Research in Astronomy, Inc., under a cooperative agreement with the NSF on behalf of the Gemini partnership: the National Science Foundation (United States), the Science and Technology Facilities Council (United Kingdom), the National Research Council (Canada), CONICYT (Chile), the Australian Research Council (Australia), Ministério da Ciência e Tecnologia (Brazil) and SECYT (Argentina).  CA acknowledges the AAS for financial support.  BR acknowledges partial support from the NASA Astrobiology Institute under Cooperative Agreement No. NNA04CC08A.

% REFERENCES 

%\clearpage 

% Figure
%\begin{figure}[tb] 
%\includegraphics*[angle=0,scale=0.75]{v1647-feb07-opt.ps} 
%\epsscale{0.8}
%\epsscale{0.9}
%\plotone{v1647ori-r-lc-2008.ps} 
%\caption{The light curve of V1647~Ori from the first observation taken after the %discovery of its brightening in February 2004 to its fade to its pre-outburst %''quiescent'' brightness in early 2007.  The data from August 2008 is shown as the %filled circle.  Data from Aspin \& Reipurth (2009) is shown as open circles and that %from Ojha et al. (2006) is shown as open squares.  Uncertainties are given on each data %point and are typically smaller than the symbol size.
%\label{lc2008}}
%\end{figure}
%\clearpage

\end{document}